\documentclass[twocolumn,prb,showpacs,preprintnumbers,amsmath,amssymb]{revtex4}

\usepackage{graphicx}
\usepackage{dcolumn}
\usepackage{bm}

\begin{document}

\title{Spin and orbital structure of uranium compounds \\
on the basis of a $\bm{j}$-$\bm{j}$ coupling scheme}

\author{Takashi Hotta}

\affiliation{Advanced Science Research Center,
Japan Atomic Energy Research Institute,
Tokai, Ibaraki 319-1195, Japan}

\date{\today}

\begin{abstract}
Key role of orbital degree of freedom to understand the magnetic structure
of uranium compounds is discussed from a microscopic viewpoint
by focusing on typical examples such as UMGa$_5$ (M=Ni and Pt) and
the mother compound UGa$_3$.
By analyzing an orbital degenerate Hubbard model constructed from
the $j$-$j$ coupling scheme in the cubic system, we obtain
the phase diagram including several kinds of magnetic states.
Especially, in the parameter region corresponding to actual uranium
compounds, the phase diagram suggests successive transitions among
paramagnetic, magnetic metallic, and insulating N\'eel states, consistent
with the experimental results for AuCu$_3$-type uranium compounds.
Furthermore, taking account of tetragonal effects such as
level splitting and reduction of hopping amplitude along the $z$-axis,
an orbital-based scenario is proposed to understand the change
in the magnetic structure from G- to A-type antiferromagnetic phases,
experimentally observed in UNiGa$_5$ and UPtGa$_5$.
\end{abstract}



\maketitle

\section{Introduction}

Recently, $f$-electron compounds with the HoCoGa$_5$-type
``115'' tetragonal crystal structure
have been intensively investigated in both the experimental and
theoretical research fields of condensed matter physics.
Among such compounds, interesting magnetic properties have been reported
for UMGa$_5$, where M is a transition metal ion.
\cite{U115-1,U115-2,U115-3,U115-4,Tokiwa,U115-5,U115-6,U115-7,U115-8}
In particular, neutron scattering experiments have revealed that
UNiGa$_5$ exhibits the G-type antiferromagnetic (AF) phase,
while UPdGa$_5$ and UPtGa$_5$ have the A-type AF state.
\cite{Tokiwa,U115-8}
Note that G-type indicates a three-dimensional N\'eel state,
while A-type denotes a layered AF structure in which
spins align ferromagnetically (FM) in the $ab$ plane and
AF along the $c$ axis.\cite{Wollan}
It is quite interesting that the magnetic structure is different
for U-115 compounds which differ only by the substitution
of transition metal ions.

U-115 materials have also been found to differ in their
magnetic anisotropy.
For M=Ni, Pd, and Pt, both $\chi_a$ and $\chi_c$ exhibit
Curie-Weiss behavior, but $\chi_a$ is somewhat larger
than $\chi_c$,\cite{U115-8}
where $\chi_a$ and $\chi_c$ are magnetic susceptibilities
for magnetic field parallel to the $a$ and $c$ axes, respectively.
This anisotropy increases in the sequence 
UNiGa$_5$, UPdGa$_5$, and UPtGa$_5$.
On the other hand, for M=Co, Rh, Ir, Fe, Ru, and Os,
both $\chi_a$ and $\chi_c$ are almost independent of temperature,
since these are Pauli paramagnets,
but in these cases $\chi_c$ is somewhat larger than $\chi_a$.
Nonetheless, we again observe a tendency for the anisotropy to become larger
for the progression from $3d$ to $4d$ and then to $5d$  M ions.
Thus, it is characteristic of U-115 compounds that
the magnetic properties are sensitive to the choice of transition
metal ions.

Furthermore, UGa$_3$, which is the mother compound of UMGa$_5$,
has also provided intriguing experimental results.
It has been reported that UGa$_3$ exhibits a G-type AF metallic phase
in the low-temperature region,\cite{UGa3}
but a ``hidden'' ordering different from the magnetic one has been
suggested by resonant X-ray scattering measurements.\cite{Mannix}
Unfortunately, orbital ordering in UGa$_3$ is not yet confirmed
experimentally, but it may be an interesting possibility
to understand the result of resonant X-ray scattering experiment
on UGa$_3$ based on the orbital-ordering scenario.

Here we note that one must pay close attention to the meanings
of ``spin'' and ``orbital'' in $f$-electron systems.
Since they are tightly coupled with each other through
a strong spin-orbit interaction, distinguishing them is not
straightforward in comparison with $d$-electron systems.
This point can create serious problems when we attempt to
understand microscopic aspects of magnetism and
superconductivity in $f$-electron compounds.
Thus, it is necessary to carefully define the terms ``orbital''
and ``spin'' for $f$ electrons in a microscopic discussion of
magnetism and superconductivity in uranium compounds.
In order to overcome this difficulty, we have proposed to employ
a $j$-$j$ coupling scheme to discuss $f$-electron systems.\cite{Hotta}

Let us note the advantages of the $j$-$j$ coupling scheme at the outset.
First, it is quite convenient for the inclusion of many-body
effects using standard
quantum-field theoretical techniques, since individual $f$-electron states
are clearly defined.
In contrast, in the $LS$ coupling scheme we cannot use such standard
techniques, since Wick's theorem does not hold.
Second we can, in principle, include the effects of valence fluctuations.
In some uranium compounds, the valence of the uranium ion is neither
definitely U$^{3+}$ nor U$^{4+}$, indicating that the $f$-electron
number takes a value between 2 and 3.
In the $j$-$j$ coupling scheme this is simply regarded
as the average number of $f$ electron per uranium ion.

In this paper, then, we propose a new scenario based on the $j$-$j$ coupling
scheme in order to understand the magnetic properties of uranium compounds,
of which UGa$_3$ and UMGa$_5$ are considered to be typical examples.
A microscopic model constructed from the $j$-$j$ coupling scheme is
analyzed using an unbiased method, such as the Lanczos technique.
For UGa$_3$ we obtain a phase diagram, including
the paramagnetic (PM) state and several types of AF phases.
In particular, we note that the PM state is adjacent to an AF metallic
phase, consistent with the experimental results for uranium
compounds having the AuCu$_3$-type crystal structure.
We also discuss the orbital structure of the AF metallic phase.
By taking account of the effects of two-dimensionality, we find that
the change in magnetic structure from G- to A-type AF phases
is reproduced, consistent with experimental results for UMGa$_5$.
Moreover, it is shown that the observed trends of the magnetic anisotropy
within the UMGa$_5$ series can be understood from the present results.

The organization of this paper is as follows.
In Sec.~II, we will introduce a microscopic model Hamiltonian
formulated on the basis of the $j$-$j$ coupling scheme.
In Sec.~III, numerical results for UGa$_3$ and UMGa$_5$ are
discussed in detail and compared with experimental results.
In Sec.~IV, future developments are discussed and the paper is
summarized.
In the Appendix, we analyze the local $f$-electron configuration
in order to examine the validity of the $j$-$j$ coupling picture
$vis$-$a$-$vis$ experimental results on the level scheme in certain
$f$-electron materials.
Throughout the paper, we use units such that $\hbar$=$k_{\rm B}$=1.

%
%
\section{Model Hamiltonian}

In this section we derive a microscopic Hamiltonian for UGa$_3$
and UMGa$_5$ based on the $j$-$j$ coupling scheme.
Although it is difficult to determine the valence of the uranium ion,
here we assume that the formal valence is U$^{3+}$,
including three $f$ electrons per ion.
By considering the crystalline electric field (CEF) potential
and Coulomb interactions, we then assign three electrons to states 
in the $j$=5/2 sextet.

\subsection{Local $f$-electron states}

First, we define the one $f$-electron states
in the AuCu$_3$-type cubic crystal structure.
The effects of tetragonality will be discussed later.
From the work of Hutchings for the case of cubic symmetry,
\cite{Hutchings}
we identify two eigen energies as
$-240B_4^0$ for the $\Gamma_7$ doublet
and $120B_4^0$ for the $\Gamma_8$ quartet,
where $B_4^0$ is a cubic CEF parameter.

In order to proceed with the discussion, it is necessary to
know which is lower, $\Gamma_7$ or $\Gamma_8$,
in the one $f$-electron picture.
For some crystal structures it is possible to determine
the level scheme from intuitive discussions of
$f$-electron wavefunctions and the positions of ligand ions.
However, this is not the case for the AuCu$_3$-type crystal structure.
For this case we invoke certain experimental results
on CeIn$_3$, a typical AuCu$_3$-type Ce-based compound,
where $\Gamma_7$ and $\Gamma_8$ have been reported as ground and excited
states, respectively, with an energy difference of 12meV.\cite{Knafo}
Thus, we take $\Gamma_7$ to be lower for the present considerations,
as shown in Fig.~1(a).

In the $j$-$j$ coupling scheme for UGa$_3$, we accommodate
three electrons in the one-electron energy states
$\Gamma_7$ and $\Gamma_8$.
There are two possibilities, i.e., low- and high-spin states,
depending on the Hund's rule interaction
and the splitting between the $\Gamma_7$ and $\Gamma_8$ levels.
Noting that the effective Hund's rule interaction is small
in the $j$-$j$ coupling scheme,\cite{Hotta}
the low-spin state should be realized, as shown in Fig.~1(b).
In fact, this low-spin state is also consistent with 
the $LS$ coupling scheme.
Details regarding electron configurations in the $j$-$j$ coupling
scheme are discussed in the Appendix.

In the electron configuration shown in Fig.~1(b), the
$\Gamma_7$ level is fully occupied to form a singlet.
If this $\Gamma_7$ level is located well below the $\Gamma_8$,
the occupying electrons will not contribute to the magnetic properties.
Thus, we may be allowed to ignore the $\Gamma_7$ electrons
for our present purposes, but this simplification
will be examined more carefully in the following subsection.

\begin{figure}[t]
\includegraphics[width=0.5\linewidth]{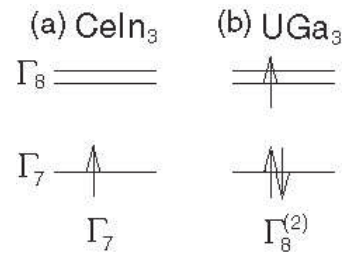}
\caption{Configurations of $f$ electrons in the $j$-$j$ coupling
scheme for (a) CeIn$_3$ and (b) UGa$_3$.
Note that up and down arrows denote pseudospin states in order
to distinguish the two states of the Kramers doublet.}
\end{figure}

\subsection{Suppression of $\Gamma_7$}

First, we discuss the $f$-electron kinetic term given by
\begin{equation}
 H_{\rm kin}=
 -\sum_{{\bf i,a},\tau,\tau',\sigma}t^{\bf a}_{\tau\tau'}
 f^{\dag}_{{\bf i}\tau\sigma}f_{{\bf i+a}\tau\sigma},
\end{equation}
where $f_{{\bf i}\tau\sigma}$ is the annihilation operator
for an $f$-electron with pseudospin $\sigma$ in the $\tau$-orbital
at site ${\bf i}$, and
$t^{\bf a}_{\tau\tau'}$ is the $f$-electron hopping matrix element
between $\tau$-and $\tau'$-orbitals along the ${\bf a}$ direction.
Indices $a$ and $b$ denote the $\Gamma_8^{\rm a}$ and
$\Gamma_8^{\rm b}$ orbitals, respectively, while $c$ indicates
the $\Gamma_7$.
Here, the pseudospin has been introduced to distinguish
the two states of the Kramers doublet.

When we evaluate $t^{\bf a}_{\tau\tau'}$ for
nearest-neighbor hopping via the $\sigma$ bond, it is given by
\begin{equation}
  t_{\tau\tau'}^{\bf x} = t
\left(
\begin{array}{ccc}
3/4 & -\sqrt{3}/4 & 0 \\
-\sqrt{3}/4 & 1/4 & 0 \\
0 & 0 & 0 \\
\end{array}
\right),
\end{equation}
for the $x$-direction,
\begin{equation}
  t_{\tau\tau'}^{\bf y} = t
\left(
\begin{array}{ccc}
3/4 & \sqrt{3}/4 & 0 \\
\sqrt{3}/4 & 1/4 & 0 \\
0 & 0 & 0 \\
\end{array}
\right),
\end{equation}
for the $y$-direction, and
\begin{equation}
  t_{\tau\tau'}^{\bf z} = 
\left(
\begin{array}{ccc}
0 & 0 & 0 \\
0 & t_z & 0 \\
0 & 0 & 0 \\
\end{array}
\right),
\end{equation}
for the $z$-direction. Note that $t_z$=$t$ in the cubic system.
In this paper, $t$ is taken as the unit of energy.
Later, $t_z$ will be treated as an extra parameter
in order to take account of the tetragonality.
We immediately notice that
non-zero hoppings occur only among $\Gamma_8$ orbitals.
Since the $\Gamma_7$ orbital has nodes along the cubic axes,
it is localized in the present tight-binding approximation.
This further justifies the suppression of $\Gamma_7$.

One may question the foregoing approach on the grounds that
it is based entirely on nearest-neighbor hopping processes.
In order to address this concern, it is useful to examine the
results of band-structure calculations, e.g.,
for CeIn$_3$ (Ref.~\onlinecite{Betsuyaku})
and UGa$_3$ (Ref.~\onlinecite{Harima}).
Note that both results have been obtained assuming
the system is in the paramagnetic state.
In order to focus on the $f$ electron components of
the energy band,
we concentrate on the bands around the $\Gamma$ point
near the Fermi level.
For CeIn$_3$, the energy band dominated by $\Gamma_7$ character
is found to be lower than the $\Gamma_8$-dominated band,
consistent with the local level scheme in Fig.~1(a).
An important point is that the Fermi level intersects the
$\Gamma_7$-dominant band, indicating that the Fermi surface
is mainly composed of $\Gamma_7$ electrons hybridized with
Ga-ion $p$ electrons. 

On the other hand, for UGa$_3$, the $\Gamma_7$ band
is also lower than the $\Gamma_8$ band, but here
the Fermi level crosses the $\Gamma_8$ band.
Thus, the $\Gamma_7$ band appears to be fully occupied,
consistent with the $j$-$j$ coupling level scheme, as
shown in Fig.~1(b).
Since the main contribution to the Fermi surface
comes from $\Gamma_8$ electrons, it is natural to dwell 
on the $\Gamma_8$ bands and ignore the occupied $\Gamma_7$ bands
in giving further consideration to many-body effects.

It is useful to consider the Fermi-surface structure of $H_{\rm kin}$
in comparison with that of the band calculations.
In Fig.~2, we show Fermi-surface sheets derived from $H_{\rm kin}$
for the case of $\langle n \rangle $=1, where $\langle n \rangle$
indicates the average number of $\Gamma_8$ electrons per site.
Owing to the multi-orbital nature of the problem, we observe two
Fermi-surface sheets.
A cube-like Fermi surface is centered on the $\Gamma$ point and
a second sheet is composed of three connected tubes
along the three cubic axes.

In the band-structure calculation results, two Fermi surface sheets
are also observed.\cite{Harima}
One sheet surrounds the $\Gamma$ point with a small hole at the center.
Another is a large sphere-like Fermi surface centered at the R point.
Since the carrier number is simply fixed at unity per site
in the tight-binding model, it is difficult to obtain perfect agreement.
Except for details, however, $H_{\rm kin}$ can reproduce
the Fermi-surface structure of the band calculation
in spite of its simplification.

\begin{figure}[t]
\includegraphics[width=\linewidth]{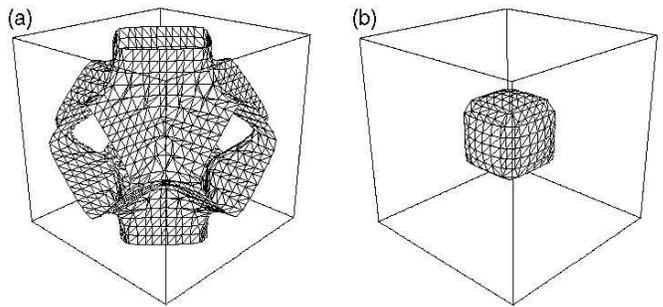}
\caption{Fermi-surface sheets of the $\Gamma_8$ tight-binding model
for $\langle n \rangle$=1.
Note that we obtain two Fermi-surface sheets, (a) and (b),
due to the multi-orbital nature of the model.
Note also that both sheets are depicted in the electron picture.
The bounding box indicates the first Brillouin zone for a simple cubic
lattice. The $\Gamma$ point is located at the center of the box,
while the apices denote R points.}
\end{figure}

\subsection{$\Gamma_8$ model}

Following the above discussion, the effective Hamiltonian for
uranium compounds will be expressed by a $\Gamma_8$-orbital
degenerate Hubbard model, given as
\begin{equation}
   H = H_{\rm kin} + H_{\rm CEF} + H_{\rm int},
\end{equation}
where $H_{\rm kin}$ denotes the $\Gamma_8$ part of Eq.~(1) and
$H_{\rm CEF}$ is written as
\begin{equation}
   H_{\rm CEF}=-\Delta\sum_{\bf i}(n_{{\bf i}a}-n_{{\bf i}b}).
\end{equation}
Here $n_{{\bf i}\tau}$=$\sum_{\sigma}n_{{\bf i}\tau\sigma}$
and $n_{{\bf i}\tau\sigma}$=
$f_{{\bf i}\tau\sigma}^{\dag} f_{{\bf i}\tau\sigma}$.
The level splitting $\Delta$ between $\Gamma_8$ orbitals is
introduced in order to consider certain tetragonal CEF effects  
for the case of UMGa$_5$.
When we analyze the magnetic properties of UGa$_3$,
this term is not needed.
For the time being, $\Delta$ is then set to be zero.

The Coulomb interaction term $H_{\rm int}$ is given by
\begin{eqnarray}
 H_{\rm int} &=&
 U \sum_{{\bf i},\tau}n_{{\bf i}\tau\uparrow}n_{{\bf i}\tau\downarrow}
 +U'\sum_{\bf i}n_{{\bf i}a} n_{{\bf i}b}
 \nonumber \\
 &+& J/2 \sum_{{\bf i},\sigma,\sigma'}\sum_{\tau \ne \tau'} 
 f_{{\bf i}\tau\sigma}^{\dag} f_{{\bf i}\tau'\sigma'}^{\dag}
 f_{{\bf i}\tau\sigma'} f_{{\bf i}\tau'\sigma} \nonumber \\
 &+& J' \sum_{\bf i} \sum_{\tau \ne \tau'} 
 f_{{\bf i}\tau\uparrow}^{\dag} f_{{\bf i}\tau\downarrow}^{\dag}
 f_{{\bf i}\tau'\downarrow} f_{{\bf i}\tau'\uparrow},
\end{eqnarray}
where $U$, $U'$, $J$, and $J'$ denote
intra-orbital, inter-orbital, exchange, and pair-hopping
interactions, respectively.
These are expressed in terms of Racah parameters, and
the relation $U$=$U'$+$J$+$J'$ holds due to the rotational
invariance in orbital space.\cite{Hotta}
For $d$-electron systems, one also has the relation $J$=$J'$.
When the electronic wavefunction is real, this relation is
easily demonstrated from the definition of the Coulomb integral.
However, in the $j$-$j$ coupling scheme the wavefunction is
complex, and $J$ is not equal to $J'$ in general.
For simplicity, we shall assume here that $J$=$J'$,
noting that essential results are not affected.
Since double occupancy of the same orbital is suppressed
owing to the large value of $U$, pair-hopping processes
are irrelevant in the present case.

%
%
\section{Results}

Among several possible methods to analyze the present microscopic model,
in this paper we have employed the exact diagonalization technique.
Although this has the drawback that it is difficult to enlarge the
system size, this method offers the clear advantage that it is possible to
deduce the magnetic structure in an unbiased manner.
In order to discuss the ground-state properties,
it is useful to calculate both the spin and orbital correlations,
which are defined, respectively, by
\begin{equation}
  S(\bm{q}) = (1/N)\sum_{{\bf i,j}}
  \langle \sigma^{z}_{\bf i} \sigma^{z}_{\bf j} \rangle
  e^{i\bm{q}\cdot({\bf i}-{\bf j})},
\end{equation}
with $\sigma^{z}_{\bf i}$=
$\sum_{\tau}(n_{{\bf i}\tau\uparrow}-n_{{\bf i}\tau\downarrow})/2$
and
\begin{equation}
  T(\bm{q}) = (1/N)\sum_{{\bf i,j}}
  \langle {\bf \tau}^{z}_{\bf i}{\bf \tau}^{z}_{\bf j} \rangle
  e^{i\bm{q}\cdot({\bf i}-{\bf j})},
\end{equation}
with $\tau^{z}_{\bf i}$=$(n_{{\bf i}a}-n_{{\bf i}b})/2$.
Here $N$ is the number of sites.

\subsection{Cubic system}

First let us consider the cubic system.
Due to severe limitations of computer memory,
our calculations are restricted to a 2$\times$2$\times$2 cube.
Nonetheless, essential points on the spin structure can be captured.
In Fig.~3(a), as a typical result for spin correlation,
we show $S(\bm{q})$ as a function of $J$ for $U'$=6.
From the changes in the dominant component,
we can define three regions as (I) $J$$\alt$0.45,
(II) 0.45$\alt$$J$$\alt$2.05, and (III) $J$$\agt$2.05.
In region I, the dominant component of $S(\bm{q})$
appears at $\bm{q}$=$(\pi,\pi,\pi)$, indicating a G-type
AF structure.
Definitions of spin structure are shown in Fig.~3(b).
Note that for C-type
[$\bm{q}$=$(\pi,\pi,0)$, $(\pi,\pi,0)$, $(\pi,\pi,0)$]
and A-type
[$\bm{q}$=$(\pi,0,0)$, $(0,\pi,0)$, $(0,0,\pi)$] structure,
$S(\bm{q})$ has the same value for each $\bm{q}$
due to the cubic symmetry.
In region II, 0.45$\alt$$J$$\alt$2.05,
the cubic symmetry seems to be broken.
Of course, this is spurious because of the smallness of the system,
but for reasons given below, this phase is considered to be
``metallic'', and it is conventionally called the PM phase.
Note that the AF correlation with $\bm{q}=(\pi,\pi,\pi)$ is still
dominant for 0.45$\alt$$J$$\alt$0.8, indicating that the system
is in a metallic magnetic phase.
For 0.8$\alt$$J$$\alt$2.05, A-type AF correlations turn out to
be dominant.

\begin{figure}[t]
\includegraphics[width=1.0\linewidth]{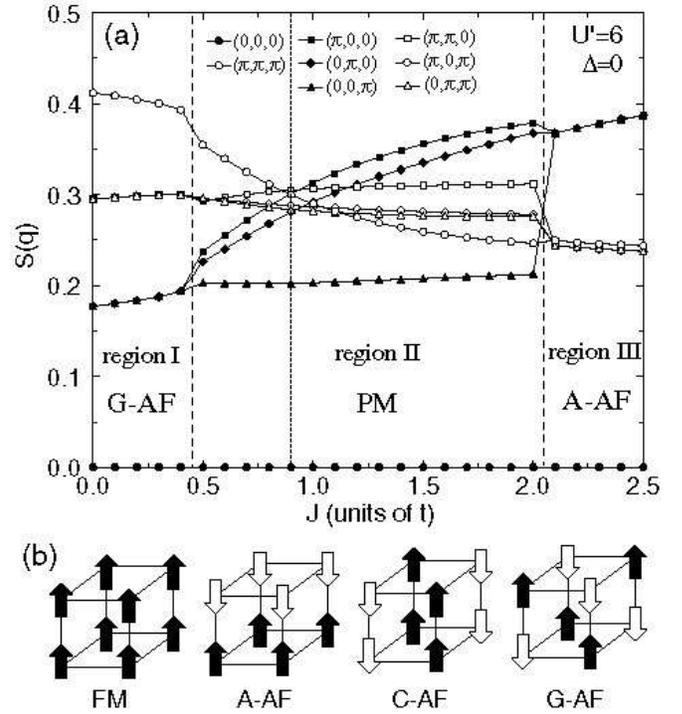}
\caption{(a) Spin correlation $S(\bm{q})$
as a function of $J$ for $U'$=6.
(b) Spin structures in the FM, A-type AF, C-type AF,
and G-type AF phases.}
\end{figure}

Here we discuss the evidence for a crossover between
insulating and metallic behavior given by the spin correlation function.
In an insulating phase, the spin structure is essentially
determined by a round trip for an electron just between
neighboring sites,
leading to orbital-dependent superexchange interactions.
Thus, cubic symmetry is maintained in the spin correlation
function even for a small-sized cluster
such as a 2$\times$2$\times$2 cube.
However, in the PM metallic phase, electrons tend to gain
kinetic energy by moving around the whole system.
Such a motion depends sensitively on the anisotropy of
the shape of the orbital.
Specifically, in a small-sized cluster,
the spin correlation function in the PM phase depends
on the choice of the basis set for orbitals,
and the cubic symmetry of the spin correlations
appear to be broken owing to the smallness of the system.
Thus, the spurious violation of cubic symmetry
in the spin correlations is considered to be a signal for a metallic phase.
Note that the metal-insulator boundary itself depends
on the system size.
In the thermodynamic limit, cubic symmetry should exist
in the metallic phase, since the effect of orbital anisotropy is
smeared by averaging over the whole system.
In fact, the band dispersion relation has cubic symmetry,
as is easily checked by diagonalizing $H_{\rm kin}$
in momentum space.

When we further increase the value of $J$, the spin
correlations recover cubic symmetry in region III, $J$$\agt$2.05.
Again we find an insulating phase, but the spin structure is now
considered to be A-type AF.
The change from G- to A-type AF phase with increasing $J$
can be understood in terms of the competition between kinetic and magnetic
energies, in analogy with manganites.\cite{Hotta2}
Namely, for large $J$ there is an energy gain for an FM spin pair
on neighboring sites, but not for an AF spin pair.
Thus, there is a tendency for the occurrence of ferromagnetism
in the large-$J$ region.
When we compare the magnetic energy of nearest-neighbor pairs,
we notice that the number of FM pairs in the A-type AF phase is
larger than for the G-type phase.
Thus, for large values of $J$ the A-type phase appears.

\begin{figure}[t]
\includegraphics[width=0.9\linewidth]{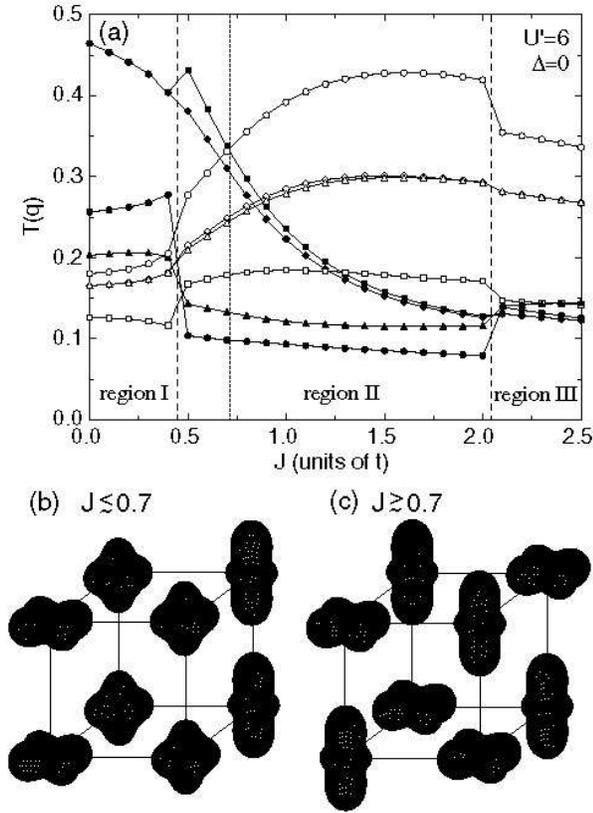}
\caption{(a) Orbital correlation as a function of $J$ for $U'=6$.
Meanings of the symbols are the same as those in Fig.~3(a).
Orbital arrangements (b) for $J$$\alt$0.7 and (c) for $J$$\agt$0.7.}
\end{figure}

In Fig.~4(a) we show typical results for orbital correlations
as a function of $J$ for $U'$=6.
Corresponding to the changes in the dominant component of
the spin correlation, we again see three regions.
In region I ($J$$\alt$0.45), $T(\bm{q})$ has
two degenerate dominant components of
$\bm{q}$=$(\pi,0,0)$ and $(0,\pi,0)$.
The orbital pattern is given by a mixture of two A-type orbital
correlations, as schematically shown in Fig.~4(b).
Since the Hamiltonian is not invariant under orbital exchange,
$T(\bm{q})$ should not have the same magnitude for
$\bm{q}$=$(\pi,0,0)$, $(0,\pi,0)$, and $(0,0,\pi)$.
On the other hand, in  region III ($J$$\agt$2.05) the
$(\pi, \pi, \pi)$ component is dominant,
suggesting a G-type orbital pattern as shown in Fig.~4(c).

\begin{figure}[t]
\includegraphics[width=0.8\linewidth]{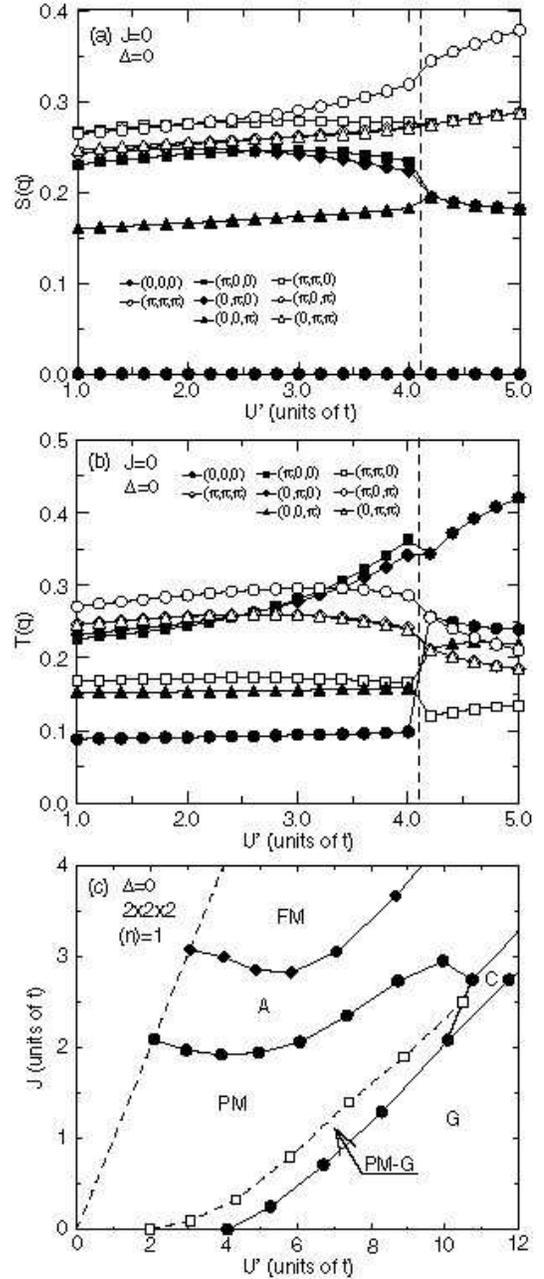}
\caption{(a) Spin correlation as a function of $U'$ for $J$=0.
(b) Orbital correlation as a function of $U'$ for $J$=0.
(c) Phase diagram for UGa$_3$ obtained by the exact
diagonalization.
The region of $J$$>$$U'$ is ignored, since it is unphysical.
See Fig.~3(b) for the definitions of abbreviations.
Here ``PM-G'' indicates the PM phase with
enhanced $(\pi, \pi, \pi)$ spin correlation.}
\end{figure}

Now we focus our attention to the region II,
in which we can see the crossover
in the orbital pattern between Figs.~4(b) and (c).
Note that the crossover point is different from that between
G- and A-type AF phases observed in $S(\bm{q})$.
This may be related to the appearance of two distinct
transition temperatures of UGa$_3$ around at $T_{\rm N}$.\cite{Mannix}
Namely, one is related to ordering of spin and another is
originating from orbital degree of freedom.
An interesting point is that both orbital patterns,
Figs.~4(b) and (c), are antiferro-like in the basal plane.
This may be detected in future by careful experiments.

In order to consider the spin and orbital structure of
uranium compounds, let us focus on the small-$J$ region.
For $J$=0, we show $S(\bm{q})$ as a function of $U'$ in Fig.~5(a).
In the PM region for small $U'$ such as for $U'$$\alt$2,
there is no dominant component in $S(\bm{q})$, but for $U'$$\agt$2,
$(\pi,\pi,\pi)$ correlation gradually grows. 
The PM phase with dominant $(\pi, \pi, \pi)$
spin correlation is  defined as ``PM-G'' in this paper.
With further increasing $U'$, eventually the ground state becomes
G-type AF insulating for $U'$$\agt$4.

In the experimental results for UX$_3$, when the lattice constant
becomes large in the order of X=Si, Ge, Sn, and Pb, the system is
changed from Pauli paramagnetic to AF metallic.\cite{Onuki}
For X=Al, Ga, In, and Tl, a similar change has been reported.
\cite{Aoki,Haga}
If we simply consider that the increase of the lattice constant
leads to the decrease in the effective hopping,
$U'/t$ becomes large with the increase of the lattice constant.
Then, the present phase diagram in the small-$J$ region
is consistent with the experimental results for UX$_3$.

In Fig.~5(b), the orbital correlation as a function of $U'$
is shown For $J$=0.
In the region corresponding to PM-G, both orbital patterns,
Figs.~4(b) and (c), are possible in principle,
but we note that the pattern (b) appears for larger value of $U'$
in the PM-G region near the G-AF insulating phase.
Since UGa$_3$ is an AF metal, it may be natural
to consider that the orbital correlation corresponding to Fig.~4(b)
can be detected in actual compounds.
Note also that two types of orbital arrangement, Figs.~4(b) and (c),
should be distinguished by the structure along the $c$-axis.
This point may be clarified in future experiments.

After we have performed calculations for several parameter sets,
the ground-state phase diagram is completed on the $(U', J)$ plane,
as shown in Fig.~5(c).
The PM phase exists for large parameter space and
in the boundary region between PM and G-type AF states,
we can see the PM phase with dominant $(\pi, \pi, \pi)$ spin
correlation.
Note that such a PM-G region is not specific to the case of $J$=0,
since it appears even when we increase the Hund's rule interaction.

Here we briefly discuss the phases in the large-$J$ region.
We observe an interesting similarity with the phase diagram for
undoped manganites RMnO$_3$,\cite{Hotta3}
in which mobile $e_{\rm g}$-electrons are tightly coupled with
the Jahn-Teller distortions and the background $t_{\rm 2g}$ spins.
Note that the present Hamiltonian is just equal to the $e_{\rm g}$
electron part of the model for manganites.\cite{Hotta2}
In the so-called double-exchange system with large Hund's rule
coupling between $e_{\rm g}$ and $t_{\rm 2g}$ electrons,
the Jahn-Teller distortion suppresses the probability of
double occupancy and it plays a similar role as
the interorbital Coulomb interaction $U'$.
The AF coupling among $t_{\rm 2g}$ spins, $J_{\rm AF}$,
controls the FM tendency in the $e_{\rm g}$-electron phases.
Roughly speaking, large (small) $J_{\rm AF}$ denotes small
(large) $J$.
Then, we see an interesting similarity between Fig.~5(c) and
the phase diagram for manganites, except for the PM region.
Especially, a chain of the transition,
FM$\rightarrow$A-AF$\rightarrow$C-AF$\rightarrow$G-AF,
occurs with decreasing $J$ (increasing $J_{\rm AF}$).
Again we stress that the present $\Gamma_8$ model for
$f$-electron systems is essentially the same as
the $e_{\rm g}$ orbital model in the $d$-electron systems.
It is interesting to observe common phenomena
concerning orbital degree of freedom in $f$-electron systems.

\subsection{Tetragonal system}

In the previous subsection, we have analyzed our model
for the case of a cubic system in an effort to understand UGa$_3$.
Our analysis has yielded a magnetic, metallic phase with
antiferro-like orbital correlations, consistent with
the experimental results.
In order to extend this discussion to tetragonal systems such as
UMGa$_5$, it is necessary to introduce two new ingredients into
the model Hamiltonian.

One is a non-zero value for $\Delta$, which is a 
level splitting between two of the orbitals.
Under a tetragonal CEF, there are 
two $\Gamma_7$ levels and one $\Gamma_6$.
Of these the $\Gamma_6$ state is the same as $\Gamma_8^{b}$
for the cubic system.
The two $\Gamma_7$ states are given by linear combinations of
the $J_{z}$=$\pm3/2$ and $\mp5/2$ states,
which can also be expressed as an admixture of
$\Gamma_7$ and $\Gamma_8^{a}$.
For simplicity, we introduce $\Delta$ as a splitting energy
between the $\Gamma_8$ orbitals.

Secondly, it is necessary to change the hopping amplitude
along the $z$-axis.
In UMGa$_5$, the MGa$_2$ layer is sandwiched between two UGa$_3$ sheets,
suggesting that the hopping amplitude of $f$-electrons
along the $z$-axis should be reduced from that in UGa$_3$.
It is difficult to estimate this reduction quantitatively,
since one must include correctly the hybridization with
$d$-electrons in the transition metal ions and with $p$-electrons
in the Ga ions. Thus, we manage the anticipated reduction
by simply treating $t_z$ as a parameter.

In Figs.~6(a) and (b), typical results for spin and orbital correlations
are shown for $t_{z}$=0.8, $J$=0, and $U'$=3.5.
For $|\Delta|$$\gg$1, a G-AF phase is observed, since the Hamiltonian is
effectively reduced to a single-band model at half-filling, and
the superexchange interaction stabilizes the G-AF state.
However, the A-AF phase appears for $-0.39$$\alt$$\Delta$$\alt$$-0.17$
near the orbitally degenerate region.
The mechanism of the appearance of the A-AF phase in the negative $\Delta$
region will be discussed later, in terms of an orbital-based scenario.
Regarding the orbital structure, for $|\Delta|$$\gg$1,
simple ferro-orbital (FO) ordered phases are obtained.
In the narrow region where $-0.23$$\alt$$\Delta$$\alt$$-0.05$,
we have identified an antiferro orbital (AFO) pattern,
as shown in Fig.~4(c).
Although the crossover point from an FO to an AFO pattern deviates slightly
from the A-AF and G-AF phase boundary, it is basically considered that
the A-AF phase appears in the region with an FO pattern.

In Fig.~6(c), we show the phase diagram in the
$(\Delta, t_z)$ plane for $J$=0 and $U'$=3.5,
in which the ground state for $\Delta$=0 and $t_z$=1
is magnetic metallic, as seen in Fig.~5(c).
It is found that an A-type AF phase appears in the negative
$\Delta$ region for $t_z$$\agt$0.68.
Note that the appearance of the A-AF phase is not sensitive to $t_z$
as long as $t_z$$\agt$0.68.
Rather, $\Delta$ seems to play a key role in controlling the change of
the magnetic phase. Here we recall the experimental fact that
UNiGa$_5$ exhibits a G-type AF phase,
while UPtGa$_5$ shows an A-type.\cite{Tokiwa}
Thus, it is necessary to relate the effect of $\Delta$ to
the difference in magnetic structure
found between UNiGa$_5$ and UPtGa$_5$.
Although $t_z$ may differ among U-115 compounds,
we focus here on the effect of $\Delta$.

\begin{figure}[t]
\includegraphics[width=0.9\linewidth]{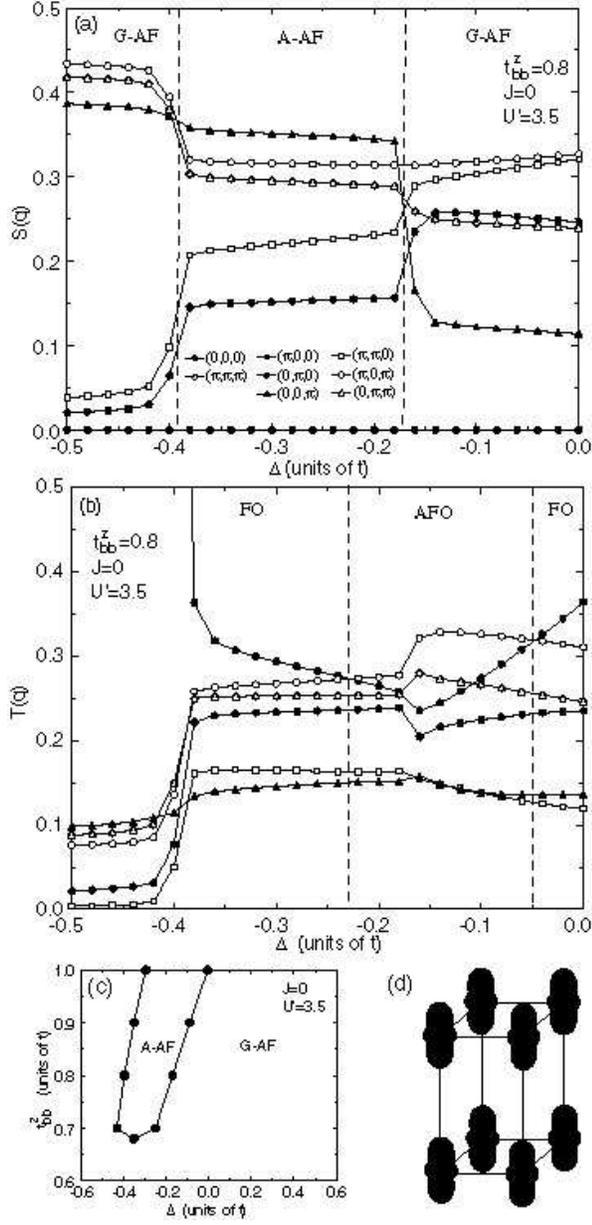}
\caption{Spin (a) and orbital (b) correlation functions vs.
$\Delta$ for $t_z$=0.8, with $J$=0 and $U'$=3.5.
(c) Phase diagram of the magnetic structure
in the $(\Delta, t_z)$ plane for $J$=0 and $U'=3.5$.
(d) Ferro orbital pattern in the A-type AF phase.}
\end{figure}

Let us now discuss the reasons for the appearance of an A-AF phase.
For negative values of $\Delta$, we easily obtain FO ordering
composed of $\Gamma_8^{b}$ orbitals, as illustrated in Fig.~6(d).
For electrons to gain kinetic energy of motion along the $z$-axis,
it is necessary to place the AF spin arrangement along this same axis.
In the FM spin configuration, electrons cannot move along the $z$-axis
due to the Pauli principle, since hopping occurs $only$ between
$\Gamma_8^{b}$ orbitals along the $z$-axis.
On the other hand, in the $xy$ plane $b$-orbital electrons
can hop to neighboring $a$-orbitals with a significant amplitude,
which is larger than that between neighboring $b$-orbitals.
Thus, in order to gain kinetic energy, electrons tend to occupy
$a$-orbitals even in the FO state composed of $b$-orbitals,
as long as $|\Delta|$ is not so large.
When we explicitly include the effects of the Hund's rule
interaction $J$, electron spins should have FM alignment between
neighboring sites in order to gain energy
in hopping processes from $b$- to $a$-orbitals.
Consequently, a FM spin configuration is favored in the $xy$ plane.
In fact, in spite of the FO state for $-0.39$$<$$\Delta$$<$0,
we can see a significant component of $T(\pi,\pi,\pi)$.
In cases with antiferro orbital correlations,
spin correlation tends in general to be FM,
as has been widely recognized in orbitally degenerate systems.

Here we mention a relation of $\Delta$ to
the magnetic anisotropy in U-115 materials.
For UPtGa$_5$ with the A-AF phase, $\chi_a$ is larger than $\chi_c$,
whereas this anisotropy is not pronounced in UNiGa$_5$
with the G-AF phase.\cite{U115-8}
An analysis for the high-temperature region based on $LS$ coupling
yields the $J_z$=$\pm$1/2 Kramers doublet as the ground state
among the dectet of $J$=9/2 ($L$=6 and $S$=3/2).\cite{Walstedt}
The states with $J_z$=$\pm$1/2 in the $LS$ coupling scheme
have significant overlap with
$f^{\dag}_{{\bf i}b \uparrow}f^{\dag}_{{\bf i}c \uparrow}
f^{\dag}_{{\bf i}c \downarrow}|0 \rangle$ and
$f^{\dag}_{{\bf i}b \downarrow}f^{\dag}_{{\bf i}c \uparrow}
f^{\dag}_{{\bf i}c \downarrow}|0 \rangle$
in the $j$-$j$ coupling scheme.
Accordingly, by the present definition $\Delta$ should be
negative to place $\Gamma_8^{b}$ below $\Gamma_8^{a}$.
If the absolute value of $\Delta$($<$0) becomes large,
$\Gamma_8^{b}$ is well separated from $\Gamma_8^{a}$
and the magnetic anisotropy will consequently become large.
Thus, a change from G- to A-type AF phase is consistent with
the trends of magnetic anisotropy in UNiGa$_5$ and UPtGa$_5$.


Finally, we make a brief comment about the effect of $t_{z}$.
Following the above discussion, the A-AF phase should
appear even for small $t_{z}$.  However, in the persent
calculation it disappears for $t_{z}$$\alt$0.68,
a critical value which seems to be rather large.
Such a quantitative point depends on the system size, and
we note that it is necessary to perform the calculation
in the thermodynamic limit.
This is a problem for future consideration.

%
%
\section{Discussion and Summary}

In this paper, we have proposed a calculational model with
active orbital degrees of freedom in an effort to understand
the magnetic structure of uranium compounds from a
microscopic point of view.
In order to construct such a model, we have incorporated
the $j$-$j$ coupling scheme, in which
one-electron states are defined first, and
Coulomb interactions are included subsequently.
This approach is consistent with the itinerant picture for
$f$ electrons.
By using an exact diagonalization technique,
we have found a magnetic metallic phase with antiferro-like
orbital correlations for UGa$_3$ and a change in the magnetic
structure from G- to A-type AF phases for UMGa$_5$.

In an effort to understand the magnetism and superconductivity
of $f$-electron systems from a microscopic standpoint,
we have carried out a study of an orbitally degenerate model.
While such investigations are just beginning, we already see
a number of opportunities for future work along this path.
Concerning issues directly related to the present paper,
it is highly recommended that calculations be carried out
in the thermodynamic limit, in order
to confirm the present exact diagonalization results.
For instance, the magnetic susceptibility should be evaluated
in the random phase approximation or fluctuation-exchange method.
With such an approach, the magnetic structure can be discussed by detecting
the divergence in the magnetic susceptibility.
This is one of our future tasks.
Another problem is how to establish the effective reduction of
$t_{z}$ in considering the case of UMGa$_5$.
In such systems, MGa$_2$ sheets are interspersed between UGa$_3$ layers,
but the main process may occur through the Ga ions.\cite{Harima}
To analyse this, it is necessary to treat
a three-dimensional $f$-$p$ model with
explicit consideration of U and Ga ions.
This is another problem for future investigation.

We comment briefly on the CEF levels of 115 compounds.
In this paper, we have considered the UMGa$_5$ systems by introducing
a splitting energy $\Delta$ between $\Gamma_8$ orbitals.
This procedure did not include changes in the basis wavefunctions.
Nonetheless, in following the present strategy,
a better way to consider U-115 might be to accommodate
three electrons in the CEF levels of Ce-115.
Regarding the latter systems, there are some interesting
results in the literature.
In analyses of the susceptibility and specific heat,
\cite{Takeuchi,Shishido} for example,
the ground level has been considered to be $\Gamma_7^{(2)}$,
consistent with $\chi_a$$<$$\chi_c$ in Ce-115,
and the first excited level has been suggested to be
$\Gamma_7^{(1)}$, not $\Gamma_6$.
Regarding CeRhIn$_5$, neutron scattering experiments have
been performed in spite of the inclusion of indium,
\cite{Christiansen} identifying energy levels in the 
sequence $\Gamma_7^{(2)}$, $\Gamma_7^{(1)}$, and $\Gamma_6$,
in agreement with previous analyses.\cite{Shishido}
If we simply put three electrons into the CEF levels
of CeRhIn$_5$, assuming Hund's rule coupling to be small,
it may be difficult to understand the appearance of
A-type AF phases, since $\Delta$ becomes positive
within the level scheme of Ce-115.

A simple way to understand such a discrepancy is to remark
that Ce-115 contains In and M = Co, Ir, and Rh,
while the U-115 considered here contains Ga and M = Ni, Pd, and Pt.
Different ions lead to different effects on the level scheme.
In fact, the magnetic anisotropy of UCoGa$_5$ (Pauli paramagnet)
is small compared with that of UNiGa$_5$, and $\chi_c$ is slightly larger
than $\chi_a$.
In UFeGa$_5$, which is also a Pauli paramagnet, we have found that
$\chi_a$$<$$\chi_c$, opposite to what is observed
in UNiGa$_5$ and UPtGa$_5$.
We envisage that in UMGa$_5$, $\Delta$ is positive for M=Fe,
slightly positive for M=Co, and negative for M=Ni.
Moreover, $|\Delta|$ increases in the sequence $3d$, $4d$,
and $5d$ transition metal ions, consistent with the magnetic
anisotropy.

In summary, we have analyzed an orbitally degenerate model appropriate
for UGa$_3$ and UMGa$_5$ using an exact diagonalization technique.
What we have found is a magnetic metallic phase with antiferro-like
orbital correlations.
By introducing tetragonal effects such as level splitting and
a reduced hopping amplitude along $z$-axis,
we have reproduced the change in the spin structure from G- to
A-type AF phases, corresponding to UNiGa$_5$ and UPtGa$_5$.

\section*{Acknowledgement}

The author thanks H. Harima for fruitful discussions
on the band-structure calculation results for CeIn$_3$ and UGa$_3$.
He is also grateful to R. E. Walstedt for valuable comments and
critical reading of the manuscript.
Stimulating discussions with Y. Haga, S. Ikeda, S. Kambe, K. Kaneko,
H. Kato, T. Maehira, T. D. Matsuda, N. Metoki, H. Onishi, Y. \=Onuki,
T. Takimoto, and K. Ueda have benefited the present paper.
This work has been supported by the Grant-in-Aid for Scientific
Research from Japan Society for the Promotion of Science.
The computation in this work has been partly done using the facilities
of the Supercomputer Center, Institute for Solid State Physics,
University of Tokyo

%
%
\appendix

\section{Level scheme}

In this Appendix, we examine a local $f$-electron state
based on the $j$-$j$ coupling scheme, comparing it with
that obtained from the $LS$ coupling scheme and from experimental results.
Concerning the level scheme for the $f^1$ electron state,
we consider two cases.

First, we consider the AuCu$_3$-type cubic crystal structure
with one $f$ electron per site.
As already mentioned in the main text, a typical material is
CeIn$_3$, in which $\Gamma_7$ and $\Gamma_8$ are the ground and
first excited states, respectively,\cite{Knafo}
as shown in Fig.~7(a).
If we accommodate one more electron to consider the $f^2$ configuration,
immediately there appear two possibilities,
``low'' and ``high''-spin states.
When the CEF splitting energy between $\Gamma_7$ and
$\Gamma_8$ levels is smaller than the Hund's rule coupling,
the second electron should be accommodated in the $\Gamma_8$ levels.
In the situation in which one is in the $\Gamma_7$ and the other
in the $\Gamma_8$, a $\Gamma_4$ or $\Gamma_5$ triplet appears
for the $f^2$ state in general, but under special conditions,
a $\Gamma_3$ doublet can occur.
On the other hand, if the CEF splitting is larger than
the Hund's rule interaction, then the $f^2$ ground state is formed
from two $\Gamma_7$ electrons, leading to a $\Gamma_1$ state.
When we compare this $\Gamma_1$ state with that in
the $LS$ coupling scheme, we notice that it is given by
a mixture of $J$=0 and $J$=4 states,
but the $J$=4 component is found to be dominant.
Note also that $\Gamma_1$ is the antisymmetric representation of
$\Gamma_7 \times \Gamma_7$.

Since we do not know the exact value of the Hund's rule
interaction in $f$-electron compounds, it is difficult to
determine the $f^2$ state by purely theoretical arguments.
In this case, we have to refer to data on actual materials.
Fortunately, we have the example of PrIn$_3$, a typical $f^2$ material
with AuCu$_3$-type crystal structure.
From several experimental results, $\Gamma_1$ has been
confirmed to be the ground level in PrIn$_3$.\cite{PrIn3}
Thus, the low-spin state should be taken for the AuCu$_3$-type
structure in the $j$-$j$ coupling scheme.

\begin{figure}[t]
\includegraphics[width=0.8\linewidth]{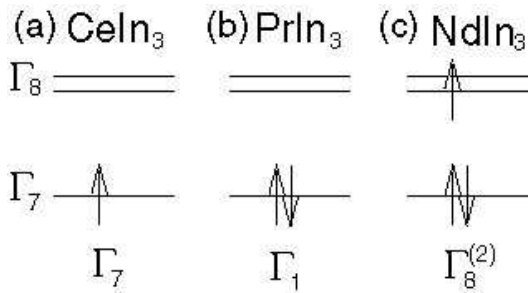}
\caption{Electron configurations in the $j$-$j$ coupling
scheme for AuCu$_3$-type compounds,
(a) CeIn$_3$, (b) PrIn$_3$, and (c) NdIn$_3$.}
\end{figure}

Here the reader may pose a naive question:
Is the Hund's rule interaction really that small in $f$-electron systems?
We have already had to answer this question in a previous paper.\cite{Hotta}
In a word, we are considering the effective Hund's rule interaction
in the $j$-$j$ coupling scheme, not in the $LS$ coupling scheme.
The original form for the Hund's rule interaction is
written as $-J_{\rm H} \bm{s}_{\bf i}^2$,
where $\bm{s}_{\bf i}$ denotes the operator for the
``real'' $f$-electron spin at site ${\bf i}$
and $J_{\rm H}$ is the Hund's rule interaction among $f$ electrons
in $\ell$=3 orbitals.
In order to transfer this to the $j$-$j$ coupling scheme,
it is convenient to use the well-known relation
${\bm s}_{\bf i}$=$(g_J$$-1){\bm j}_{\bf i}$,
where $g_J$ is the Land\'e's $g$-factor and ${\bm j}_{\bf i}$
is the operator for the total angular momentum of $j$=5/2
at site ${\bf i}$.
From standard textbooks, we easily obtain $g_J$=6/7,
indicating that ${\bm s}_{\bf i}$=$-(1/7){\bm j}_{\bf i}$.
Thus, the Hund's rule term in the $j$-$j$ coupling scheme is
rewritten as $-J_{\rm eff} \bm{j}_{\bf i}^2$ with
$J_{\rm eff}$=$J_{\rm H}/49$.
Note then that the magnitude of the Hund's rule interaction is
effectively reduced by the factor 1/49 in the $j$-$j$ coupling scheme.
Even if $J_{\rm H}$=1eV, $J_{\rm eff}$ is reduced to be about 200K,
which is comparable with the CEF splitting energy.
Thus, it is possible to have the low-spin state in the $j$-$j$
couping scheme.

Next, we take a further step to the $f^3$ state by adding one more
$f$ electron.
Since $\Gamma_7$ is fully occupied to form $\Gamma_1$,
the next electron should be placed in the $\Gamma_8$ state
as shown in Fig.~7(c), clearly indicating that there exists
an active orbital degree of freedom.
The $f^3$ state composed of two $\Gamma_7$ and one $\Gamma_8$
electron is expressed as $\Gamma_8^{(2)}$ in the terminology of
group theory.
When we again consider actual materials,
NdIn$_3$ is found to be a typical $f^3$ material with
the AuCu$_3$-type crystal structure.
In experiments, it has been established
that $\Gamma_8^{(2)}$ is the ground level,\cite{NdIn3}
as we have found with the present $j$-$j$ coupling scheme.

Let us turn our attention to another crystal structure in which
$\Gamma_8$ is lower than $\Gamma_7$ in the $f^1$ configuration.
Typical materials are the rare-earth hexaborides RB$_6$
with R=Ce, Pr, and Nd.
As is well known, the ground level of CeB$_6$ is $\Gamma_8$,
indicating that the quadrupolar degree of freedom plays an
active role in this material.\cite{CeB6}
In fact, anomalous behavior related to quadrupolar ordering
has been suggested by several exeprimental results.

First, we note that the level splitting between $\Gamma_8$ and
$\Gamma_7$ is assumed to be larger than the Hund's rule interaction.
When we accommodate two electrons in $\Gamma_8$ orbitals,
the triplet ($\Gamma_5$), doublet ($\Gamma_3$), and
singlet ($\Gamma_1$) states are allowed.
Among these, owing to the effect of the Hund's rule interaction,
even if it is small, the $\Gamma_5$ triplet should be
the ground state. 
This has actually been observed in PrB$_6$.\cite{Loewenhaupt,PrB6}
Further, in order to consider NdB$_6$,
yet another electron is put into the $\Gamma_8$ orbital,
making a total of three.
Alternatively, we may say that there is one hole in the $\Gamma_8$ orbital.
Such a state is found, again, to be characterized by $\Gamma_8^{(2)}$.
Experimental results on NdB$_6$ have actually been reported which lead
to the ground state of $\Gamma_8^{(2)}$.\cite{Loewenhaupt,NdB6}
Thus, when $\Gamma_8$ is the ground state for the one $f$-electron
case, we obtain $\Gamma_5$ for the $f^2$ and $\Gamma_8^{(2)}$
for the $f^3$ configurations.

\begin{figure}[t]
\includegraphics[width=0.8\linewidth]{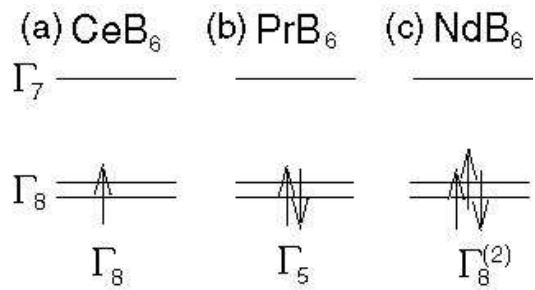}
\caption{Electron configurations in the $j$-$j$ coupling
scheme for rare-earth hexaborides,
(a) CeB$_6$, (b) PrB$_6$, and (c) NdB$_6$.}
\end{figure}

We have shown that the ground states deduced from the $j$-$j$ couping
scheme are consistent with experimental results.
However, in order to explain the experimental results quantitatively,
it is unavoidable to analyze the CEF levels
using the $LS$ coupling scheme.
What we would like to stress is that even in a localized system,
the symmetry of the ground level can be understood via the
$j$-$j$ coupling scheme.
We need to recognize the limitations of the $j$-$j$ coupling scheme
when we treat a local electronic state.
For instance, to consider the $f^3$ state, we simply put
three electrons into the CEF level scheme
which is determined with the $f^1$ configuration.
Thus, the wavefunction of the $f^3$ state is uniquely determined.
However, in an actual situation, the dectet labelled by $J$=9/2
($L$=6 and $S$=3/2) is split into two $\Gamma_8$ and one $\Gamma_6$
orbital.
The ground-state wavefunctions will then depend on the
two CEF parameters $B_4^0$ and $B_6^0$.\cite{LLW}
As mentioned above, in order to explain experimental
results on localized $f$-electron materials, ones should analyze
the CEF effects in the system using the $LS$ coupling scheme.
In this paper, however, the electronic states are considered with
an itinerant picture based on the $j$-$j$ coupling scheme.
Thus, it is important to check that the local electronic state
formed by $f$ electrons in this way is consistent with the symmetry of
the state obtained with the $LS$ coupling scheme.

In summary, it has been shown that the ground states of the $f^2$
and $f^3$ configurations can be qualitatively reproduced by
accommodating $f$ electrons in the CEF levels of a corresponding
$f^1$ material, provided that the CEF level splitting is larger
than the Hund's rule interaction.
Thus, the $j$-$j$ coupling scheme works even in the localized case.
Accordingly, we believe that a microscopic theory can be developed
in which we discuss the magnetism and superconductivity of $f$-electron
compounds in terms of the $j$-$j$ coupling scheme.


\end{document}